\documentclass[aps,prl,twocolumn,amsmath,amssymb,nofootinbib,superscriptaddress,floatfix]{revtex4-1}

\usepackage{amsmath}
\usepackage{amssymb}
\usepackage{amsthm}
\usepackage[dvips]{color} 
\usepackage{graphicx}
\usepackage{dcolumn} 
\usepackage{bm} 
\usepackage[hypertex]{hyperref}
\usepackage{longtable}
\usepackage{ulem}   
\begin{document}

\title{Virtual Parallel Computing and a Search Algorithm using Matrix
  Product States}

\author{Claudio Chamon} \affiliation{Department of Physics, Boston
  University, Boston, Massachusetts 02215, USA}

\author{Eduardo R. Mucciolo} \affiliation{Department of Physics,
  University of Central Florida, Orlando, Florida 32816, USA}

\date{\today}

\begin{abstract}
   We propose a form of parallel computing on classical computers that
   is based on matrix product states. The virtual parallelization is
   accomplished by representing bits with matrices and by evolving
   these matrices from an initial product state that encodes multiple
   inputs. Matrix evolution follows from the sequential application of
   gates, as in a logical circuit. The action by classical
   probabilistic one-bit and deterministic two-bit gates such as NAND
   are implemented in terms of matrix operations and, as opposed to
   quantum computing, it is possible to copy bits. We present a way to
   explore this method of computation to solve search problems and
   count the number of solutions. We argue that if the classical
   computational cost of testing solutions (witnesses) requires less
   than $O(n^2)$ local two-bit gates acting on $n$ bits, the search
   problem can be fully solved in subexponential time. Therefore, for
   this restricted type of search problem, the virtual parallelization
   scheme is faster than Grover's quantum algorithm.
\end{abstract}

\maketitle

Interference and the ability to follow many history paths
simultaneously make quantum systems attractive for implementing
computations \cite{deutsch85}. Efficient algorithms exploring these
properties have been proposed to solve practical problems such as
number factoring \cite{shor94} and unsorted database search
\cite{grover97}. However, we still do not have a sufficiently large
and resilient quantum computer to take advantage of these
algorithms. It is, thus, very desirable to try to find better and more
efficient ways to compute with classical systems. In this regard,
recent advances in our understanding of quantum many-body systems
provide some guidance. It is well understood now that the time
evolution of a large class of one-dimensional interacting systems can
be efficiently simulated by expressing their wave functions in a
matrix product state form and by using a time-evolving block
decimation (TEBD) \cite{vidal}. A key aspect of this success is data
compression: even though many-body interactions tend to increase the
rank of the matrices over time, it is possible to use truncation along
the evolution to keep the matrices relatively small, such that the
resulting wave function approximates quite accurately the exact one
without an exponential computation cost \cite{verstraete2008}.
In quantum systems, it is well understood that local interactions do
not quickly entangle one-dimensional many-body state, justifying the
matrix truncation \cite{cirac2009,hamma2011}.

In this Letter, we describe a method of classical computation that
utilizes matrix product states (MPS) to implement search and other
similar tasks. Compression, when possible, provides additional
speedup. Formally, instead of working with wave functions and quantum
amplitudes, we describe the state of the computer in terms of a
stochastic probability distribution written as traces of matrix
product states associated to bit configurations. The idea of
expressing classical probability distributions in the form of MPS is
not new \cite{derrida1993}, but the focus so far has been on using it
to study nonequilibrium phenomena of physical systems (see for
instance Ref. \onlinecite{temme2010,johnson2010}). As we show below, a
MPS formulation of classical probability distributions can also be
employed to create a virtual parallel machine where all possible
outcomes of an algorithm are obtained for all $2^n$ inputs of an
$n$-bit register. Information about these outcomes is encoded and
compressed in the matrices forming the MPS. By itself this
``parallelism'' is not obviously useful; it is, however, if a certain
problem can use the probability of a {\it single} outcome at a
time. This is the case of a search problem that seeks, for a given
$y$, the value of $x$ such that $y=f(x)$ for an algorithmically
computable function $f$. Then, the focus is not on all values of the
output, but on only one given $y$. We shall show below that in this
case matrix computing can be useful. In particular, from the
probability of $y$, the method directly provides the number of input
values $x$ satisfying the functional constraint $y=f(x)$.

In our matrix computing, insertion and removal of bits are allowed and
one-bit and two-bit gates can be implemented much like in a
conventional computer. Our one-bit gates are probabilistic while our
two-bit gates are deterministic. Two-bit gates rely on a singular
value decomposition (SVD) to maintain the MPS form of the probability
distribution. All these operations preserve the positivity and the
overall normalization of the probability even though we work with
nonpositive matrices.

{\it Matrix computing formulation} -- Consider a set of binary
variables $\{x_j=0,1\}_{j=1,\ldots,n}$ describing a set of $n$ bits,
with $|x_1\,x_2 \ldots x_n\rangle\equiv |x\rangle$ denoting a
particular configuration of this system. In analogy to quantum
mechanics, we define the vector
\begin{equation}
\label{eq:Psi}
|P\rangle = \sum_{x_n,\ldots,x_1=0,1} P(x_1,\ldots,x_n)\, |x_1\ldots
 x_n\rangle,
\end{equation}
where
\begin{equation}
\label{eq:P}
P(x_1,\ldots,x_n) = \mbox{tr} \left( M_1^{x_1} \cdots
M_n^{x_n} \right).
\end{equation}
Here, each $M_j^{x_j}$ is a real matrix of dimensions $D_{j-1}\times
D_{j}$. The trace can be dropped if we consider the first and last
matrices to be row and column vectors, {\it i.e.}, $D_0=D_n=1$. The
state vector is normalized in the following sense: define
$|\Sigma\rangle = \sum_{x_1,\ldots,x_n=0,1} |x_1\ldots x_n\rangle$,
then $Z = \langle \Sigma |P\rangle=1$ since $\sum_x P(x)=1$.

Starting from an initial probability distribution
$P_0(x_1,\ldots,x_n)$, the vector $|P\rangle$ evolves as a sequence of
one-bit and nearest-neighbor two-bit gates is applied to the bit
matrices. These bit operations form a logical circuit, which is
tailored according to a particular computational problem, for
instance, the algorithmic computation of a function $f(x)$. Below, we
describe how bit operations are implemented.

$\bullet$ {\it One-bit gates:} We will use probabilistic one-bit
gates, which take states $0,1$ to states $0,1$ with probabilities
$p,1-p$ and $q,1-q$:
\begin{eqnarray}
&&0\xrightarrow{\;\;p\;\;} 0 \quad {\rm or} \quad 0\xrightarrow{1-p} 1
  \nonumber \\ &&1\xrightarrow{1-q}0 \quad {\rm or} \quad
  1\xrightarrow{\;\;q\;\;} 1 \nonumber \;.
\end{eqnarray}
The probabilities can be encoded into a transfer function $t^{\tilde
  a,a}$ that takes a logic input $a=0,1$ into a logic output $\tilde
a=0,1$. Explicitly: $t^{0,0}=p$, $t^{1,0}=1-p$, $t^{0,1}=1-q$,
$t^{1,1}=q$. A one-bit gate acting on bit $j$ yields a new matrix
\begin{equation}
  \label{eq:1-bit}
\tilde M_j^{x_j} = \sum_{x'_{j}=0,1} t^{x_{j}, x'_{j}} \;M_j^{x'_j}
\;.
\end{equation}
The transfer function satisfies the sum rule $\sum_{{\tilde a}=0,1}
t^{\tilde a,a}=1$, which ensures that the normalization $Z=1$ is
maintained as the system evolves. Examples of one-bit gates are: (a)
Deterministic NOT, with $p=0$ and $q=0$, (b) RAND, with $p=1/2$ and
$q=1/2$, which randomizes the bit, (c) RST, with $p=1$ and $q=0$,
which resets the bit to 0.

$\bullet$ {\it Two-bit gates:} We will consider only deterministic
two-bit gates. Given two logical functions $A(a,b)$ and $B(a,b)$, we
construct the transfer function $T^{\tilde{a}\tilde{b},ab}$, taking
bits with states $a$ and $b$ to bits with states $\tilde{a}$ and
$\tilde{b}$, respectively:
\begin{equation}
T^{\tilde a \tilde b,ab} =
\begin{cases}
1,&\text{$\tilde a=A(a,b)$\ and\ $\tilde b=B(a,b)$}, \\ 0, &
\text{otherwise} .
\end{cases}
\end{equation}
Similarly to one-bit gates, the normalization after two-bit gates is
preserved by the sum rule $\sum_{{\tilde a},{\tilde b}=0,1} T^{\tilde
  a \tilde b,ab}=1$. The evolved matrices must satisfy
\begin{equation}
  \label{eq:2-bit}
\;\tilde M_{j-1}^{x_{j-1}}\, \tilde M_j^{x_j} =
\sum_{x'_{j-1},x'_{j}=0,1} T^{x_{j-1} x_{j},x'_{j-1},x'_{j}} \,
M_{j-1}^{x'_{j-1}}\;M_j^{x'_j},
\end{equation}
and we use the SVD to decompose the result of the gate operation on
the right-hand side of Eq.~(\ref{eq:2-bit}) as a product of two
matrices, as in the left-hand side of the equation, for all the four
cases $x_{j-1},x_{j}=0,1$.

Let us demonstrate this construction with a concrete example. Consider
the following logical operation on bits $j-1$ and $j$: $A_{\rm
  NAND}(a,b)=a$ and $B_{\rm NAND}(a,b)=\overline{a\land{}b}$. The
first bit is unaffected, while the second one evolves into the NAND
operation between the two bits. In this case, $T^{01,00} = T^{01,01} =
T^{11,10} = T^{10,11} = 1$, with all other elements set to zero. We
use the transfer function to determine the four blocks (for
$x_{j-1},x_{j}=0,1$) of a matrix ${\cal M}^{\rm NAND}_{j-1,j}$ of
dimension $2D_{j-2} \times 2D_{j}$:
\begin{equation}
\label{eq:bigM}
{\cal M}^{\rm NAND}_{j-1,j} = \left( \begin{array}{c|cc} 0 & M_{j-1}^0
  M_j^0 + M_{j-1}^0 M_j^1 \\ \\ \hline \\ M_{j-1}^1 M_j^1 & M_{j-1}^1
  M_j^0
\end{array} \right).
\end{equation}
To factor the matrix ${\cal M}_{j-1,j}$ back into a product, we employ
an SVD,
\begin{equation}
{\cal M}_{j-1,j} \stackrel{\rm SVD}{=} \left( \begin{array}{c}
  \tilde{M}_{j-1}^0 \\ \\ \tilde{M}_{j-1}^1 \end{array} \right)
\left( \begin{array}{cc} \tilde{M}_j^0 & \tilde{M}_j^1 \end{array}
\right).
\end{equation}
In this process, the common dimension $D_{j-1}$ may change and likely
increase. This is an issue of fundamental important, which we shall
return when we discuss a search algorithm.

$\bullet$ {\it Bit insertions and removals:} For computational tasks
such addition and multiplication, it is important to be able to insert
and remove bits. These operations are straightforward for
MPS. Insertion of a new bit (say, initially set to 0) in between bits
$j-1$ and $j$ amounts to replacing $M_{j-1}^{x_{j-1}} M_{j}^{x_{j}}$
with $M_{j-1}^{x_{j-1}} M_\alpha^{x_\alpha} M_{j}^{x_{j}}$, where
$M_\alpha^1$ and $M_\alpha^0$ are $D_{j-1}\times D_{j-1}$ null and
identity matrices, respectively, and the total sum over bit
configurations in the vector $|P\rangle$ [see Eq. (\ref{eq:Psi})] has
now to include the binary variable $x_\alpha=0,1$. Removal of a bit is
done by absorbing its matrix into the one of an adjacent bit, namely,
by tracing it out; for instance, we use $\sum_{x_j=0,1} M_j^{x_j}
M_{j+1}^{x_{j+1}} = \tilde M_{j+1}^{x_{j+1}}$ to remove bit $j$.

\vspace{.2cm}

{\it How can matrix computing be used to solve certain computational
  problems?} -- Here we shall present computational algorithms that
explore the virtual parallelism encoded in matrix product states. To
be concrete, consider the following search problem as an example:

\noindent
\begin{itemize}
\item[]
\noindent
Given a function $y=f(x)$ that can be computed algorithmically with
$O(n^d)$ gates and a certain value for $y$, we would like to search
for an input $x$ that yields as output $y=f(x)$.
\end{itemize}

The reason why matrix computation is useful for this search problem
can be argued as follows. Matrix product states can express the
probability values of all possible $m$-bit outputs $y\equiv
y_1\,y_2\dots y_m$ if one starts with a product state encoding all
possible $n$-bit inputs $x\equiv x_1\,x_2\dots x_n$, namely,
$P(x)=2^{-n}$ for all $x$. Of course, if we were interested in all the
probabilities, we would have to compute an exponentially large ($2^m$)
number of traces of products of matrices. But this is {\it not} what
is needed to perform the search above: we are interested in just {\it
  one} output $y$ for this problem. We, thus, proceed in the following
steps.

\begin{enumerate}
\item Starting with all bits $x_i$, $i=1,\dots ,n$, randomized with
  equal probabilities $1/2$ for being $0$ or $1$, we compute the final
  output matrices $M_j^{y_j}$, $j=1,\ldots,m$, resulting from the
  action of the circuit that evaluates $f(x)$.

\item We compute the probability $P(y)$ for the given $y$ we are
  interested in. If $P(y)\geq 2^{-n}$, then there is at least one
  value of $x$ such that $y=f(x)$.

\item We then fix one of the input bits, say $x_1$, to be $0$, instead
  of randomizing it. We recompute the output matrices $M_j^{y_j}$,
  $j=1,\ldots,m$, and the new probability $P(y)$. Again we test if the
  probability is above the threshold, $P(y)\geq 2^{-n+1}$ in this
  case. If the probability fell below the threshold, we must reset
  $x_1$ to $1$. (Notice that since there may be more than one $x$ for
  a given $y$, that $P(y)$ stays above threshold does not mean that
  switching to $x_1=1$ is necessarily forbidden, but we shall stick
  instead to $x_1=0$ in this case to avoid unnecessary iterations.)

\item We repeat step 3 fixing now input bit $x_2$, then repeat it
  again fixing input bit $x_3$, and so on until we finally fix input
  bit $x_n$. At the end of $n$ steps, having fixed all the $n$ bits of
  the input, we have arrived at one value for $x$ such that $y=f(x)$.

\end{enumerate}

Let us discuss the computational cost of such algorithm. To simplify
the discussion, let us present it in terms of the largest matrix
dimension $D$ in the computations, which we shall relate to the number
$n_g$ of gates involved in the computation of the function $f(x)$. All
SVD steps involve matrices with rank smaller or equal to $D$;
therefore, the cost associate to gate operations is no more than
$O(n_g\times D^3)$. One has also to compute the trace of the matrix
products for a fixed $y$ to yield the probability $P(y)$, and this
takes time no more than $O(n\times D^3)$.  We then have to repeat the
procedure fixing bit-by-bit the $x_i$, $i=1,\dots ,n$. Therefore, in
the worst case it takes a time $O(n\times \mbox{max}\{n_g,n\} \times
D^3)$ to find $x$.

The largest computational cost comes from the SVD and trace steps,
which depend on the rank $D$ of the matrices. The crucial issue is how
$D$ scales with either the number of bits $n$ or the number of gates
$n_g$ for a given algorithm to compute $f(x)$. We shall prove below
the following result: the maximum dimension of any matrix in a
computation using $n_g$ nearest-neighbor gates in a system with $n$
bits is bounded by $D \le D_{\rm max}(n,n_g) = \min \left(
2^{\lfloor\sqrt{2 n_g}\rfloor}, 2^{\lfloor n/2\rfloor} \right)$. The
consequence of this result on the computational time is as follows. As
we argued above, the search algorithm takes a time $O(n\times
\mbox{max}\{n,n_g\}\times D^3)$. For a function $y = f(x)$ that can be
computed with $n_g\sim n^d$ gates, the time to search for an $x$ that
gives a fixed $y$ has two different behaviors depending on whether
$d<2$ or $d\ge 2$. If $d < 2$, $D_{\rm max} \sim 2^{\sqrt{2}
  \,n^{d/2}}$, and thus the search takes, in the worst possible case,
a time $O(n^{d+1} \times 2^{3\sqrt{2}\, n^{d/2}})$ using matrix
computing algorithms. If instead $d \ge 2$, $D_{\rm max}$ saturates to
$D_{\rm max} \sim 2^{n/2}$ and in the worst possible case the
computation (without discarding singular values) takes exponential
time. In other words, there is a transition between subexponential and
exponential behavior at $d_c = 2$. It, thus, follows that for any
function $f(x)$ that can be computed with $n_g < O(n^2)$ gates, the
full search problem can be solved faster using matrix computing than
using Grover's quantum algorithm, which scales as $O(2^{n/2})$.

{\it Proof of the bound on the largest bond dimension} -- Upon
application of a two-bit gate on bits $j-1$ and $j$, the dimension
$D_{j-1}$ will increase as follows. Starting with $D_{j-2}\times
D_{j-1}$ matrices $M^{x_{j-1}}_{j-1}$ and $D_{j-1}\times D_{j}$
matrices $M^{x_{j}}_{j}$, one assembles a $2D_{j-2}\times 2D_{j}$
matrix ${\cal M}^{\rm gate}_{j-1,j}$ [see the example of the NAND gate
  in Eq.~(\ref{eq:bigM})]. The SVD step will lead to $D_{j-2}\times
\tilde D_{j-1}$ matrices $\tilde M^{x_{j-1}}_{j-1}$ and $\tilde
D_{j-1}\times D_{j}$ matrices $\tilde M^{x_{j}}_{j}$, where the new
bond dimension $\tilde D_{j-1}=\min(2 D_{j-2},2 D_{j})$. It is useful
to work on a logarithmic scale and define $h_j=\log_2 D_j$. Thus, we
can write $\tilde h_{j-1}=\min(h_{j-2},h_{j})+1$.

\begin{figure}[t]
\centering \includegraphics[width=8cm]{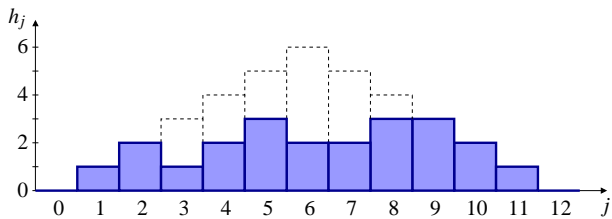}
 \caption{Example of a configuration of entanglement heights ($h_j =
   \log_2 D_j$) satisfying the height difference constraint
   $|h_{j}-h_{j-1}|\le 1, \forall j$ when $n=12$. The dashed line
   shows the configuration with maximum heights.}
\label{fig-entropy}
\end{figure}

Let us next prove that at {\it any} step in the algorithmic evolution
the ``entanglement heights'' $h_j$ satisfy the condition
$|h_{j}-h_{j-1}|\le 1, \forall j$, which we shall refer to as the
height difference constraint (hdc). The proof is done by induction. At
the initial state of the calculation, one starts with the product
state of all possible equally weighted inputs $x$, which correspond to
$1\times 1$ matrices or, equivalently, all $h_j=0$, so that
$|h_{j}-h_{j-1}|=0\le 1$, thus satisfying the condition. Now suppose
that the condition is satisfied at step $\tau$; we can show that it is
then also satisfied at step $\tau+1$, when a two-bit gate is applied
between two adjacent bits $j-1$ and $j$. None of the heights other
than $h_{j-1}\to \tilde h_{j-1}$ are changed, therefore, the hdc
condition $|h_{j}-h_{j-1}|\le 1$ remains satisfied for all $i< j-1$
and $i> j$, and it just remains to be shown that it is satisfied for
$i=j-1$ and $i=j$. Consider the case where $h_{j-2}\le h_{j}$ (the
other case $h_{j}\le h_{j-2}$ is analogous). In this case $\tilde
h_{j-1}=h_{j-2}+1$, satisfying the condition $|\tilde
h_{j-1}-h_{j-2}|\le 1$. Now $h_{j}-\tilde h_{j-1}=h_{j}-
h_{j-2}-1=(h_{j}- h_{j-1})+(h_{j-1}- h_{j-2})-1$, and using that
$h_{j}- h_{j-1}\le 1$ and $h_{j-1}- h_{j-2}\le 1$, as well as that
$h_{j-2}\le h_{j}$, we have that $|h_{j}-\tilde h_{j-1}|\le 1$. It,
thus, follows that the hdc condition $|h_{j}-h_{j-1}|\le 1, \forall j$
is satisfied at {\it all} steps in the calculation. An example of a
configuration of entanglement heights satisfying the hdc is show in
Fig. \ref{fig-entropy}.

If all we do to evolve the state is to apply two-bit gates, we have
shown that $|h_{j}-h_{j-1}|\le 1, \forall j$. It is easy to see that
after a bit insertion the condition is still satisfied, because the
change in height is zero on the two sides of the inserted bit
(corresponding to a square matrix), with all other relative height
differences unchanged. The removal (tracing out) of bits is slightly
more subtle. Right after the removal, there are large jumps across the
region where the bits were removed, but these can be brought up to
satisfy the hdc by applying a series of two-bit identity gates
[$A(a,b)=a$ and $B(a,b)=b$] sweeping from left-to-right followed by
another from right-to-left. These sweeps remove the height ``faults''
(and actually tend to decrease the overall height). Therefore we
arrive at the result that the hdc condition is satisfied after all
operations, two-bit gates, bit insertions, and bit deletions (after
the identity sweeps).

Let us now show that the maximum height resulting from the application
of $n_g$ two-bit gates is bounded by $h_{\rm max}\le \lfloor
\sqrt{2\,n_g} \rfloor$. The application of a single two-bit gate on
bits $j-1$ and $j$ changes the height $h_{j-1}\to \tilde
h_{j-1}=\min(h_{j-2},h_{j})+1$. Because the relative heights of
neighboring bonds cannot differ by more than 1 unit due to the hdc,
the maximum amount that the height $\tilde h_{j-1}$ can increase with
respect to $h_{j-1}$ is by 2 (which occurs when
$h_{j-2}=h_{j}=h_{j-1}+1$). Therefore one can write that $S=\sum_i
h_i\le 2\,n_g$. Now, suppose that the maximum height is $h_{\rm max}$
at some bond labelled by $i_{\rm max}$ (located to the right of bit
$i_{\rm max}$); because the heights $h_0$ to the left of the 1st bit
and $h_{n}$ to the right of the $n$th bit are both equal to 0 at all
times, and because of the hdc condition, there are constraints on how
quickly the heights can grow from 0 to $h_{\rm max}$ at $i_{\rm max}$
and then decrease down to 0 again. The climb and descent that minimize
the area $S$ can be trivially seen to be a triangle where $h_j$
increases linearly from $j=i_{\rm max}-h_{\rm max}$ to $j=i_{\rm
  max}$, and then decreases linearly until $j=i_{\rm max}+h_{\rm
  max}$. The area of this triangle is $S_{\rm min}=h_{\rm max}^2$, and
any other height profile that reaches the same maximum height $h_{\rm
  max}$ has equal or larger area. Therefore, $h_{\rm max}^2\le S\le
2\,n_g$, and thus we arrive at the conclusion that $h_{\rm max}\le
\lfloor \sqrt{2\,n_g} \rfloor$, {\it i.e.}, the bound on the maximum
entanglement height for a given number of gates. Furthermore, because
of the hdc and the fact that $h_0=h_{n}=0$, the entanglement height
for a fixed $j$ is bounded by $h_j\le \min(j,n-j)$, and the overall
maximum $h_{\rm max}=\lfloor n/2 \rfloor$ is reached at the center of
the chain, $j=\lfloor n/2 \rfloor$ and $j=\lceil n/2 \rceil$ (which
coincide when $n$ is even).

Putting all the conditions together, we arrive at $h_{\rm max}\le
\min\left(\lfloor \sqrt{2\,n_g} \rfloor,\lfloor n/2 \rfloor\right)$,
or equivalently, the bound $D\le D_{\rm
  max}(n,n_g)=\min\left(2^{\lfloor \sqrt{2 n_g} \rfloor},2^{\lfloor
  n/2\rfloor}\right)$ which we used to obtain the absolute maximum
running time of the search algorithm.

{\it Conclusions} -- We have shown that it is possible to achieve
virtual parallelization in single-processor classical computers using
one-bit and two-bit local gates acting on matrix product states over
$n$ bits. Based on this method, we propose a search algorithm that
runs in subexponential time when the cost to check a witness requires
less than $O(n^2)$ two-bit gates.  This critical bound in the circuit
size was obtained assuming a worst-case scenario for the matrix
dimension growth as a function of the number of two-bit
gates. However, for particular circuits, the actual rank of the
matrices may grow slower than this estimate, in which case some
speedup is possible.  In addition, during gate operations and matrix
decompositions, if the singular values decay sufficiently fast, it may
be possible to reduce matrix rank growth through truncation, similarly
to the standard procedure used in quantum methods such as the TEBD
\cite{vidal} and its classical version for stochastic evolution, the
cTEBD \cite{temme2010,johnson2010}. This question is open to future
investigation.

The method is not limited to one-dimensional bit arrays and could, in
principle, be extended to higher dimension tensor products. Finally,
we point out that the method also naturally counts the number of
satisfying assignments of a given Boolean formula, which is a problem
of much importance in computer science.

This work was supported in part by the NSF grants CCF-1116590 and
CCF-1117241.
The authors thank P. Wocjan for useful discussions.

\end{document}